\documentclass[aps,preprint,amsmath,amssymb,floatfix]{revtex4}
\usepackage{graphicx}
\begin{document}

\title{Search for anomalous quartic $WWZ\gamma$ couplings at the future linear $e^{+}e^{-}$ collider}

\author{M. K\"{o}ksal}
\email[]{mkoksal@cumhuriyet.edu.tr} \affiliation{Department of
Optical Engineering, Cumhuriyet University, 58140, Sivas, Turkey}

\author{A. Senol}
\email[]{senol_a@ibu.edu.tr} \affiliation{Department of Physics,
Abant Izzet Baysal University, 14280, Bolu, Turkey}

\begin{abstract}
In this paper, the potentials of two different processes
$e^{+}e^{-}\rightarrow W^{-} W^{+}\gamma$ and $e^{+}e^{-}
\rightarrow e^{+}\gamma^{*} e^{-} \rightarrow e^{+} W^{-} Z \nu_{e}$
at the  Compact Linear Collider (CLIC) are examined to probe the
anomalous quartic $WWZ\gamma$ gauge couplings. For $\sqrt{s}=0.5,
1.5$ and 3 TeV energies at the CLIC, $95\%$ confidence level limits
on the anomalous coupling parameters defining the dimension-six
operators are found via the effective Lagrangian approach in a model
independent way. The best limits on the anomalous couplings
$\frac{k_{0}^{W}}{\Lambda^{2}}$, $\frac{k_{c}^{W}}{\Lambda^{2}}$,
$\frac{k_{2}^{m}}{\Lambda^{2}}$ and $\frac{a_{n}}{\Lambda^{2}}$
which can be achieved with the integrated luminosity of
$L_{int}=590$ fb$^{-1}$ at the CLIC with $\sqrt{s}=3$ TeV are
$[-8.80;\, 8.73]\times 10^{-8}$ GeV$^{-2}$, $[-1.53; \, 1.51]\times
10^{-7}$ GeV$^{-2}$, $[-3.75; \, 3.74]\times 10^{-7}$ GeV$^{-2}$ and
$[-9.13;\,9.09]\times 10^{-7}$ GeV$^{-2}$, respectively.
\end{abstract}

\maketitle

\section{Introduction}

The Standard Model (SM) of particle physics has been demonstrated to
be quite successful until now through very important experimental
tests, particularly by the recent discovery of a new particle in the
mass region around $125$ GeV which is consistent with the SM Higgs
boson \cite{higgs1,higgs2}. However, the SM does not fully answer
some of the most fundamental questions such as the origin of mass,
the large hierarchy between electroweak and Planck scale, the strong
CP problem, and matter/antimatter asymmetry. To clarify these
questions, new physics beyond the SM is needed. A simple way to
discover new physics beyond SM is to probe anomalous gauge boson
self-interactions. In the electroweak sector of SM, gauge boson
self-interactions are completely determined by $SU_{L}(2)\times
U_{Y}(1)$ gauge invariance. Hence, the high precision measurements
of gauge boson self-interactions are extremely important in the
understanding of the gauge structure of the SM. Any deviation from
the expected values of these couplings would imply the existence of
new physics beyond the SM. Investigation of the new physics through
effective Lagrangian method is a well known approach. The origin of
this method is based on the assumption that at high energies above
the SM, there is a grander theory which reduces to the SM at lower
energies. Therefore, SM is supposed to be an effective low energy
theory in which heavy fields have been integrated out. Since this
fundamental method is independent of the details of the model, it is
occasionally called model independent analysis.

In this paper, we examine the anomalous quartic $WWZ\gamma$ gauge
boson couplings by analyzing two different processes
$e^{+}e^{-}\rightarrow W^{-} W^{+}\gamma$ and $e^{+}e^{-}
\rightarrow e^{+}\gamma^{*} e^{-} \rightarrow e^{+} W^{-} Z \nu_{e}$
at the CLIC. Genuine quartic couplings consisting of effective
operators, have different origins than anomalous trilinear gauge
boson couplings. Hence, we assume that genuine quartic gauge
couplings can be independently analyzed from the effects arosen from
any trilinear gauge couplings. In the literature, to examine genuine
quartic $WWZ\gamma$ couplings, there are usually two different
dimension-six effective quartic Lagrangians that keep custodial
$SU(2)_{c}$ symmetry and local $U(1)_{QED}$ symmetry. The first one,
CP-violating effective Lagrangian is given as the following
\cite{lag1}

\begin{eqnarray}
\textit{L}_{n}=\frac{i \pi \alpha}{4\Lambda^{2}} a_{n} \epsilon_{ijk} W_{\mu \alpha}^{(i)}W_{\nu}^{(j)} W^{(k)\alpha} F^{\mu\nu}
\end{eqnarray}
where $\alpha$ is the electroweak coupling constant, $W^{(i)}$ is
the $SU(2)_{c}$ weak isospin triplet, $F_{\mu \nu}$, which equals to
$\partial_{\mu}A_{\nu}-\partial_{\nu}A_{\mu}$, is the tensor for
electromagnetic field strength, $a_{n}$ represents the strength of
anomalous coupling and $\Lambda$ represents the energy scale of
possible new physics. The anomalous vertex generated from the above
effective Lagrangian is given in the Appendix.

Additionally we perform the notation of Ref. \cite{lhc} in the
writing of CP-conserving effective operators. There are fourteen
effective photonic operators associated with the anomalous quartic
gauge couplings (as shown in Eq. (5) of Ref. \cite{lhc}). They are
determined by fourteen independent couplings
$k_{0,c}^{w,b,m},k_{1,2,3}^{w,m}$ and $k_{1,2}^{b}$ that
parameterise the strength of the anomalous quartic gauge couplings.
These effective photonic operators can be described in terms of
independent Lorentz structures. Among them, the lowest order
effective $WW\gamma\gamma$ and $ZZ\gamma\gamma$ interactions are
expressed by four Lorentz invariant structures

\begin{eqnarray}
\textit{W}_{0}^{\gamma}=\frac{-e^{2}g^{2}}{2}F_{\mu \nu}F^{\mu \nu}
W^{+ \alpha} W_{\alpha}^{-},
\end{eqnarray}

\begin{eqnarray}
\textit{W}_{c}^{\gamma}=\frac{-e^{2}g^{2}}{4}F_{\mu \nu}F^{\mu
\alpha} (W^{+ \nu} W_{\alpha}^{-}+W^{- \nu} W_{\alpha}^{+}),
\end{eqnarray}

\begin{eqnarray}
\textit{Z}_{0}^{\gamma}=\frac{-e^{2}g^{2}}{4
\textmd{cos}^{2}\,\theta_{W}}F_{\mu \nu}F^{\mu \nu} Z^{\alpha}
Z_{\alpha},
\end{eqnarray}

\begin{eqnarray}
\textit{Z}_{c}^{\gamma}=\frac{-e^{2}g^{2}}{4
\textmd{cos}^{2}\,\theta_{W}}F_{\mu \nu}F^{\mu \alpha} Z^{\nu}
Z_{\alpha}.
\end{eqnarray}
In addition, the lowest order effective $ZZZ\gamma$ operators are
parameterized as

\begin{eqnarray}
\textit{Z}_{0}^{Z}=\frac{-e^{2}g^{2}}{2
\textmd{cos}^{2}\,\theta_{W}}F_{\mu \nu}Z^{\mu \nu} Z^{\alpha}
Z_{\alpha},
\end{eqnarray}

\begin{eqnarray}
\textit{Z}_{c}^{Z}=\frac{-e^{2}g^{2}}{2
\textmd{cos}^{2}\,\theta_{W}}F_{\mu \nu}Z^{\mu \alpha} Z^{\nu}
Z_{\alpha}.
\end{eqnarray}

There are only five basic Lorentz structures also related to
anomalous quartic $WWZ\gamma$ vertex as follows:

\begin{eqnarray}
\textit{W}_{0}^{Z}=-e^{2}g^{2} F_{\mu \nu}Z^{\mu \nu} W^{+ \alpha}
W_{\alpha}^{-},
\end{eqnarray}

\begin{eqnarray}
\textit{W}_{c}^{Z}=-\frac{e^{2}g^{2}}{2}F_{\mu \nu}Z^{\mu \alpha}
(W^{+ \nu} W_{\alpha}^{-}+W^{- \nu} W_{\alpha}^{+})
\end{eqnarray}

\begin{eqnarray}
\textit{W}_{1}^{Z}=-\frac{e g_{z}g^{2}}{2}F^{\mu \nu} (W_{\mu
\nu}^{+}W_{\alpha}^{-} Z^{\alpha}+W_{\mu \nu}^{-}W_{\alpha}^{+}
Z^{\alpha})
\end{eqnarray}

\begin{eqnarray}
\textit{W}_{2}^{Z}=-\frac{e g_{z}g^{2}}{2}F^{\mu \nu} (W_{\mu
\alpha}^{+} W^{-\alpha} Z_{\nu}+W_{\mu \alpha}^{-}W^{+ \alpha}
Z_{\nu})
\end{eqnarray}

\begin{eqnarray}
\textit{W}_{3}^{Z}=-\frac{e g_{z}g^{2}}{2}F^{\mu \nu} (W_{\mu
\alpha}^{+} W^{-}_{\nu} Z^{\alpha}+W_{\mu \alpha}^{-}W_{\nu}^{+}
Z^{\alpha})
\end{eqnarray}
with $g=e/ \textmd{sin}\,\theta_{W}$, $g_{z}=e/
\textmd{sin}\,\theta_{W} \textmd{cos}\,\theta_{W}$ and
$V_{\mu\nu}=\partial_{\mu}V_{\nu}-\partial_{\nu}V_{\mu}$ where
$V=W^{\pm},Z$. The vertex functions for the anomalous quartic
$WWZ\gamma$ couplings generated from Eqs. ($8$)-($12$) are given in
Appendix.

As a result, these fourteen effective operators can be written more
simply as the following:
\begin{eqnarray}
\textit{L}=&&\frac{k_{0}^{\gamma}}{\Lambda^2}(\textit{Z}_{0}^{\gamma}+\textit{W}_{0}^{\gamma})+\frac{k_{c}^{\gamma}}{\Lambda^2}(\textit{Z}_{c}^{\gamma}
+\textit{W}_{c}^{\gamma})+\frac{k_{1}^{\gamma}}{\Lambda^2}\textit{Z}_{0}^{\gamma} \nonumber \\
&&+\frac{k_{23}^{\gamma}}{\Lambda^2}\textit{Z}_{c}^{\gamma}+\frac{k_{0}^{Z}}{\Lambda^2}\textit{Z}_{0}^{Z}+\frac{k_{c}^{Z}}{\Lambda^2}\textit{Z}_{c}^{Z}+\sum_{i}\frac{k_{i}^{W}}{\Lambda^2}\textit{W}_{i}^{Z}, \nonumber \\
\end{eqnarray}
where
\begin{eqnarray}
k_{j}^{\gamma}=k_{j}^{w}+k_{j}^{b}+k_{j}^{m}\,\,\,\,\,\,\,\,\,\,\,\,\,\,\,\,(j=0,c,1)
\end{eqnarray}

\begin{eqnarray}
k_{23}^{\gamma}=k_{2}^{w}+k_{2}^{b}+k_{2}^{m}+k_{3}^{w}+k_{3}^{m}
\end{eqnarray}

\begin{eqnarray}
k_{0}^{Z}=\frac{\textmd{cos}\,\theta_{W}}{\textmd{sin}\,\theta_{W}}(k_{0}^{w}+k_{1}^{w})-\frac{\textmd{sin}\,\theta_{W}}{\textmd{cos}\,\theta_{W}}(k_{0}^{b}+k_{1}^{b})+(\frac{\textmd{cos}^{2}\,\theta_{W}-\textmd{sin}^{2}\,\theta_{W}}{2\textmd{cos}\,\theta_{W}\textmd{sin}\,\theta_{W}})(k_{0}^{m}+k_{1}^{m}),
\end{eqnarray}

\begin{eqnarray}
k_{c}^{Z}=\frac{\textmd{cos}\,\theta_{W}}{\textmd{sin}\,\theta_{W}}(k_{c}^{w}+k_{2}^{w}+k_{3}^{w})-\frac{\textmd{sin}\,\theta_{W}}{\textmd{cos}\,\theta_{W}}(k_{c}^{b}+k_{2}^{b})+(\frac{\textmd{cos}^{2}\,\theta_{W}-\textmd{sin}^{2}\,\theta_{W}}{2\textmd{cos}\,\theta_{W}\textmd{sin}\,\theta_{W}})(k_{c}^{m}+k_{2}^{m}+k_{3}^{m}),
\end{eqnarray}

\begin{eqnarray}
k_{0}^{W}=\frac{\textmd{cos}\,\theta_{W}}{\textmd{sin}\,\theta_{W}}k_{0}^{w}-\frac{\textmd{sin}\,\theta_{W}}{\textmd{cos}\,\theta_{W}}k_{0}^{b}+(\frac{\textmd{cos}^{2}\,\theta_{W}-\textmd{sin}^{2}\,\theta_{W}}{2\textmd{cos}\,\theta_{W}\textmd{sin}\,\theta_{W}})k_{0}^{m},
\end{eqnarray}

\begin{eqnarray}
k_{c}^{W}=\frac{\textmd{cos}\,\theta_{W}}{\textmd{sin}\,\theta_{W}}k_{c}^{w}-\frac{\textmd{sin}\,\theta_{W}}{\textmd{cos}\,\theta_{W}}k_{c}^{b}+(\frac{\textmd{cos}^{2}\,\theta_{W}-\textmd{sin}^{2}\,\theta_{W}}{2\textmd{cos}\,\theta_{W}\textmd{sin}\,\theta_{W}})k_{c}^{m},
\end{eqnarray}

\begin{eqnarray}
k_{j}^{W}=k_{j}^{w}+\frac{1}{2}k_{j}^{m}\,\,\,\,\,\,\,\,\,\,\,\,\,\,\,\,(j=1,2,3).
\end{eqnarray}

In this work, we are only interested in the
$k_{i}^{W}$($i=0,c,1,2,3$) parameters given in Eqs. ($18$)-($20$)
related to the anomalous $WWZ\gamma$ couplings. These $k_{i}^{W}$
parameters are correlated with couplings defining anomalous $WW
\gamma \gamma, ZZ\gamma \gamma$ and $ZZZ \gamma$ interactions
\cite{lhc}. Hence, we need to separate the anomalous $WWZ\gamma$
couplings from the other anomalous quartic couplings. This can be
achieved by imposing additional restrictions on $k_{i}^{j}$
parameters \cite{lhc1}. Thus, we set all $k_{i}^{j}$ parameters to
zero except $k_{2}^{m}$ and $k_{3}^{m}$ in the anomalous $WWZ\gamma$
couplings. Additionally, we require $k_{2}^{m}=-k_{3}^{m}$.
Therefore, the effective interactions can be obtained below
\begin{eqnarray}
\textit{L}_{eff}=\frac{k_{2}^{m}}{2\Lambda^{2}}(W_{2}^{Z}-W_{3}^{Z}).
\end{eqnarray}

In the literature, the $\frac{k_{2}^{m}}{\Lambda^{2}}$ couplings
describing the anomalous quartic $WWZ\gamma$ vertex are examined by
Refs. \cite{lhc,lag3,lhc1}. However, the
$\frac{k_{0}^{W}}{\Lambda^{2}}$ and $\frac{k_{c}^{W}}{\Lambda^{2}}$
couplings obtained with the aid of Eqs. ($18$)-($19$) provide the
current experimental limits related to the anomalous quartic
$WWZ\gamma$ couplings. In this paper, we analyze the limits on the
CP-conserving parameters $\frac{k_{0}^{W}}{\Lambda^{2}}$,
$\frac{k_{c}^{W}}{\Lambda^{2}}$ and the CP-violating parameter
$\frac{a_{n}}{\Lambda^{2}}$ which are the current experimental
limits on the anomalous quartic $WWZ\gamma$ gauge couplings, and
compare our limits with the phenomenological studies on
$\frac{k_{2}^{m}}{\Lambda^{2}}$.

Anomalous quartic $WWZ\gamma$ couplings at linear colliders and
their $e \gamma$ and $\gamma\gamma$ modes have been examined through
the processes $e^{+}e^{-}\rightarrow
W^{+}W^{-}Z,W^{+}W^{-}\gamma,W^{+}W^{-}(\gamma)\rightarrow 4f\gamma$
\cite{lin,linb,linc,lind,line}, $e \gamma\rightarrow W^{+}W^{-}e,
\nu_{e}W^{-}Z$ \cite{lag1,lin2} and $\gamma\gamma\rightarrow
W^{+}W^{-}Z$ \cite{lin3,lin4}. These couplings appear as
$W^{+}W^{-}e$ and  $\nu_{e}W^{-}Z$ final state productions of $e
\gamma$ collision at linear colliders. $\nu_{e} Z W^{-}$ production
is more sensitive to anomalous quartic $WWZ\gamma$ couplings with
respect to $eW^{-}W^{+}$ production \cite{lag1}. This production
isolates the anomalous $WWZ\gamma$ couplings from $W W \gamma
\gamma$ couplings. These couplings have also been investigated at
the Large Hadron Collider (LHC) via the processes $pp\rightarrow$
$W(\rightarrow j j) \gamma Z(\rightarrow\ell^{+}  \ell^{-})$
\cite{lhc} and $pp\rightarrow W(\rightarrow\ell \nu_{\ell}) \gamma
Z(\rightarrow\ell^{+} \ell^{-})$ \cite{lhc1}. Although anomalous
quartic $WWZ\gamma$ couplings  have been examined in many studies by
analyzing either CP-violating or CP-conserving effective Lagrangians
in the literature, these couplings have been investigated using two
effective Lagrangians only by Ref. \cite{lhc1}.

On the other hand, the limits on $\frac{a_{n}}{\Lambda^{2}}$
parameter of the anomalous quartic $WWZ\gamma$ couplings are
constrained at the LEP by analysing the process
$e^{+}e^{-}\rightarrow W^{+}W^{-} \gamma$ \cite{lep1,lep2,lep3}.
This reaction is sensitive to both the anomalous $W W \gamma \gamma$
and $WWZ\gamma$ couplings.

The latest results obtained by L3, OPAL and DELPHI collaborations
are given by $-0.14\,
\textmd{GeV}^{-2}<\frac{a_{n}}{\Lambda^{2}}<0.13\,
\textmd{GeV}^{-2}$, $-0.16\,
\textmd{GeV}^{-2}<\frac{a_{n}}{\Lambda^{2}}<0.15\,
\textmd{GeV}^{-2}$, and $-0.18\,
\textmd{GeV}^{-2}<\frac{a_{n}}{\Lambda^{2}}<0.14\,
\textmd{GeV}^{-2}$ at $95\%$ confidence level (C. L.), respectively.
However, the recent most restrictive experimental limits on
$\frac{k_{0}^{W}}{\Lambda^{2}}$ and $\frac{k_{c}^{W}}{\Lambda^{2}}$
parameters of the anomalous quartic $WWZ\gamma$ couplings are
determined through the process $q\overline{q}'\rightarrow W
(\rightarrow\ell\nu)Z(\rightarrow jj) \gamma$ by CMS collaboration
at the LHC \cite{s覺n覺r}. These are $-1.2\times 10^{-5}
\textmd{GeV}^{-2}<\frac{k_{0}^{W}}{\Lambda^{2}}<1\times 10^{-5}
\textmd{GeV}^{-2}$ and $-1.8\times 10^{-5}
\textmd{GeV}^{-2}<\frac{k_{c}^{W}}{\Lambda^{2}}<1.7\times 10^{-5}
 \textmd{GeV}^{-2}$ at $95\%$ C. L..

The LHC which is the current most powerful particle collider, is
able to carry out proton-proton collisions at $\sqrt{s}=14$ TeV. It
may generate large massive particles and allow us to reveal new
physics effects beyond the SM. However, the analysis of the LHC data
is quite difficult due to backgrounds from strong interactions. The
linear $e^{-}e^{+}$ colliders generally provide clean environment
with reference to hadron colliders and they can be used to determine
new physics effects with high precision measurements. The Compact
Linear Collider (CLIC) is one of the most popular linear colliders,
planned to realize $e^{-}$-$e^{+}$ collisions in three energy stages
of $0.5$, $1.5$, and $3$ TeV \cite{clic}. The CLIC's first energy
stage will provide an opportunity for the achievement of high
precision measurements of various observables of the SM gauge
bosons, top quark and Higgs boson. The second energy stage will
allow the detection of theories that lie beyond the SM. Moreover,
Higgs boson properties such as the Higgs self-coupling and rare
Higgs decay modes will be investigated in this stage \cite{clic1}.
CLIC's operation at $\sqrt{s}=3$ TeV reaches a higher effective
center-of-mass energy than the LHC for elementary particle
collisions \cite{clic2}. This enables the determination of new
particles and the testing of various models such as supersymmetry,
extra dimensions, and so forth beyond the LHC's capability. Besides,
the linear colliders have $e\gamma$ and $\gamma\gamma$ modes to
probe the new physics beyond the SM. High energy real photons in the
$e\gamma$ and $\gamma\gamma$ processes can be produced by converting
the original $e^{-}$ or $e^{+}$ beam into a photon beam through the
Compton back-scattering technique \cite{las1,las2}. In addition,
$e\gamma^{*}$, $\gamma \gamma^{*}$ and $\gamma^{*} \gamma^{*}$
collisions coming from quasireal photons at the linear colliders
also are examined. $e\gamma^{*}$ collision is the interaction of an
incoming lepton beam and a quasireal $\gamma^{*}$ photon associated
with the other beam particle; $\gamma \gamma^{*}$ collision is the
interaction of a real photon and a quasireal photon; and $\gamma^{*}
\gamma^{*}$ collision is the interaction between  quasireal photons.
The Weizsacker-Williams approach, known as the Equivalent Photon
Approximation (EPA), can be applied to the photons in these
processes \cite{Brodsky:1971ud, Terazawa:1973tb, es1,es2,es3}. In
the framework of EPA, the virtuality of the quasireal $\gamma^{*}$
photons is very low and they are assumed to be almost real. In EPA,
these photons carry a small transverse momentum. Hence, they deviate
at very small angles from the incoming lepton beam path. Moreover,
$e\gamma^{*}$ and $\gamma^{*} \gamma^{*}$ processes are more
realistic than $e\gamma$ and $\gamma \gamma$ processes since they
naturally occur spontaneously from the $e^{-}e^{+}$ process itself.
In the literature, photon-induced reactions through the EPA have
been extensively studied at the LEP, Tevatron, and LHC
\cite{a1,a2,a3,a4,Abazov:2010bk,Tasevsky:2011zz,a5,a6,a7,a8,a9,a10,a11,a12,a13,a14,a15,a16,a17,a18,a19,a20,a21,a22,a23,a24,a25,a26,a27}.

\section{CROSS SECTIONS AND NUMERICAL ANALYSIS }

In this work, we obtain limits on the CP-conserving parameters
$\frac{k_{0}^{W}}{\Lambda^{2}}$, $\frac{k_{c}^{W}}{\Lambda^{2}}$ and
the CP-violating parameter $\frac{a_{n}}{\Lambda^{2}}$ which are the
current experimental limits on the anomalous quartic $WWZ\gamma$
gauge couplings, and also compare our limits with phenomenological
studies on $\frac{k_{2}^{m}}{\Lambda^{2}}$ derived in Refs.
\cite{lhc,lag1,lhc1}. In order to examine our numerical
calculations, we have used the $WWZ\gamma$ vertex in CompHEP
\cite{comphep}. The general form of the total cross sections for two
processes $e^+e^-\to W^-W^+\gamma$ and $e^{+}e^{-} \rightarrow
e^{+}\gamma^{*} e^{-} \rightarrow e^{+} W^{-} Z \nu_{e}$ including
CP-conserving anomalous quartic couplings $k_i^{W}$ ($i=0,c$) can be
written as
\begin{eqnarray}\label{tcs}
\sigma_{tot}=\sigma_{SM}+\sum_i\frac{k_i^{W}}{\Lambda^2}\sigma_{int}^i+\sum_{i,j}\frac{k_i^{W}k_j^{W}}{\Lambda^4}\sigma_{ano}^{ij}
\end{eqnarray}
where $\sigma_{SM}$ is the SM cross section, $\sigma_{int}$ is the
interference terms between SM and the anomalous contribution, and
$\sigma_{ano}$ is the pure anomalous contribution. The contributions
of the interference terms to total cross section for both processes
are negligibly small comparing to pure anomalous terms. But in this
study, the small contributions of the interference terms are taken
into account in the numerical calculations. Moreover, the general
expression of the cross section including CP-violating anomalous
quartic coupling is derived by replacing $k_i^{W}=k_j^{W}$ with
$a_n$ in Eq. $(23)$. But this anomalous coupling ($a_n$) does not
interfere with the SM amplitude in all processes \cite{lag3}.
Therefore the total cross section depends only on the quadratic
function of anomalous coupling $a_n$. The total cross sections of
the process $e^{+}e^{-}\rightarrow W^{-} W^{+}\gamma$ are presented
in Figs. $1$-$4$ as functions of anomalous
$\frac{k_{0}^{W}}{\Lambda^{2}}$, $\frac{k_{c}^{W}}{\Lambda^{2}}$,
$\frac{k_{2}^{m}}{\Lambda^{2}}$ and $\frac{a_{n}}{\Lambda^{2}}$
couplings with $\sqrt{s}=0.5,1.5$ and $3$ TeV. In Figs. $1$-$4$, we
consider that only one of the anomalous quartic gauge coupling
parameters is non-zero at any given time, while the other couplings
are fixed at zero. We can see from Figs. $1$-$3$ that the value of
the anomalous cross section including
$\frac{k_{0}^{W}}{\Lambda^{2}}$ is larger than the value of
$\frac{k_{2}^{m}}{\Lambda^{2}}$ and $\frac{k_{c}^{W}}{\Lambda^{2}}$
couplings. Hence, the limits on $\frac{k_{0}^{W}}{\Lambda^{2}}$
coupling are expected to be more sensitive according to the limits
on $\frac{k_{2}^{m}}{\Lambda^{2}}$ and
$\frac{k_{c}^{W}}{\Lambda^{2}}$ couplings. Similarly, the total
cross sections of the process $e^{+}e^{-} \rightarrow
e^{+}\gamma^{*} e^{-} \rightarrow e^{+} W^{-} Z \nu_{e}$ are
presented in Figs. $5$-$8$ as functions of anomalous
$\frac{k_{0}^{W}}{\Lambda^{2}}$, $\frac{k_{c}^{W}}{\Lambda^{2}}$,
$\frac{k_{2}^{m}}{\Lambda^{2}}$ and $\frac{a_{n}}{\Lambda^{2}}$
couplings with $\sqrt{s}=0.5,1.5$ and $3$ TeV.

The $p_T$ distribution of the final state photon in $e^+e^-\to
W^-W^+\gamma$ process with the anomalous $WWZ\gamma$ couplings
$\frac{k_{0}^{W}}{\Lambda^{2}}$, $\frac{k_{c}^{W}}{\Lambda^{2}}$,
$\frac{k_{2}^{m}}{\Lambda^{2}}$ and $\frac{a_{n}}{\Lambda^{2}}$,
together with SM backgrounds at $\sqrt s$=0.5, 1.5 and 3 TeV are
given in Figs. $9$-$11$, respectively. From these figures, the final
state photon in the $e^+e^-\to W^-W^+\gamma$ process is radiated
from massless fermion-photon, $WW\gamma$ and $WWZ\gamma$ vertices.
The massless fermion-photon vertex causes infrared singularities in
the cross section. Therefore, the strong peak arises at the low
$p_T$ region of the photons. Above $p_T$ of 20 GeV we see an obvious
splitting and enhancement of the signal from SM background. The
effects of infrared singularities which diminish the contribution of
anomalous couplings to SM cross section become dominant for the high
$p_T$ region, as shown in Fig. $9$-$11$. It is clear from Fig. 9
that the distributions are more sensitive to
$\frac{k_{2}^{m}}{\Lambda^{2}}$ than to $\frac{a_{n}}{\Lambda^{2}}$.
On the other hand, at $\sqrt{s}=$ 1.5 and 3 TeV, it shows exactly
the opposite behavior. In addition, the momentum dependence of
$\frac{k_{0}^{W}}{\Lambda^{2}}$ for all center of mass energies is
bigger than $\frac{k_{c}^{W}}{\Lambda^{2}}$. Especially, the
momentum dependence of $\frac{k_{0}^{W}}{\Lambda^{2}}$  between four
different anomalous couplings is highest at $\sqrt{s}=$ 3 TeV.
Consequently, we impose a $p_T> 20$ GeV cut to reduce the SM
background without affecting the signal cross sections due to
anomalous quartic couplings.

In the course of statistical analysis, the limits of anomalous
$\frac{k_{0}^{W}}{\Lambda^{2}}$, $\frac{k_{c}^{W}}{\Lambda^{2}}$,
$\frac{k_{2}^{m}}{\Lambda^{2}}$ and $\frac{a_{n}}{\Lambda^{2}}$
couplings at $95\%$ C.L. are obtained by using $\chi^{2}$ test since
the number of SM background events of the examined processes is
greater than $10$. The $\chi^{2}$ function is defined as follows

\begin{eqnarray}
\chi^{2}=\left(\frac{\sigma_{SM}-\sigma_{NP}}{\sigma_{SM}\delta_{stat}}\right)^{2}
\end{eqnarray}
where $\sigma_{NP}$ is the total cross section in the existence of
anomalous gauge couplings, $\delta_{stat}=\frac{1}{\sqrt{N}}$ is the
statistical error in which $N$ is the number of events. The number
of expected events of the process $e^{+}e^{-}\rightarrow W^{-}
W^{+}\gamma$, $N$ is obtained by $N= L_{int} \times \sigma_{SM}
\times BR(W\rightarrow \ell \nu_{\ell})\times BR(W\rightarrow
q\bar{q}')$ where $L_{int}$ is the integrated luminosity,
$\sigma_{SM}$ is the SM cross section and $\ell=e^{-}$ or $\mu^{-}$.
Similarly, the number of expected events of the process $e^{+}e^{-}
\rightarrow e^{+}\gamma^{*} e^{-} \rightarrow e^{+} W^{-} Z \nu_{e}$
is calculated as $N= L_{int} \times \sigma_{SM} \times
BR(W\rightarrow \ell \nu_{\ell})\times BR(Z\rightarrow q \bar{q})$.
In addition, we impose the acceptance cuts on the pseudorapidity
$|\eta^{\,\gamma}|<2.5$ and the transverse momentum
$p_{T}^{\,\gamma}>20 \:$ GeV for photons in the process
$e^{+}e^{-}\rightarrow W^{-} W^{+}\gamma$. After applying these
cuts, the SM background cross sections for the process
$e^{+}e^{-}\rightarrow W^{-} W^{+}\gamma$ are $1.65\times 10^{-1}$
pb at $\sqrt{s}=0.5$ TeV, $6.00\times 10^{-2}$ pb at $\sqrt{s}=1.5$
TeV, and $2.63\times 10^{-2}$ pb at $\sqrt{s}=3$ TeV. They are
$3.58\times 10^{-3}$ pb at $\sqrt{s}=0.5$ TeV, $5.92\times 10^{-2}$
pb at $\sqrt{s}=1.5$ TeV, and $1.61\times 10^{-1}$ pb at
$\sqrt{s}=3$ TeV for the process $e^{+}e^{-} \rightarrow
e^{+}\gamma^{*} e^{-} \rightarrow e^{+} W^{-} Z e^{-}$.

The one-dimensional limits on anomalous couplings
$\frac{k_{0}^{W}}{\Lambda^{2}}$, $\frac{k_{c}^{W}}{\Lambda^{2}}$,
$\frac{k_{2}^{m}}{\Lambda^{2}}$ and $\frac{a_{n}}{\Lambda^{2}}$ at
$95\%$ C.L. sensitivity at various integrated luminosities and
center-of-mass energies are given in Tables I-VI. As can be seen in
Tables I and II, the limits on $\frac{k_{0}^{W}}{\Lambda^{2}}$,
$\frac{k_{c}^{W}}{\Lambda^{2}}$ are approximately several orders of
magnitude more restrictive than those obtained from the LHC
\cite{s覺n覺r}  while the best limits obtained on
$\frac{a_{n}}{\Lambda^{2}}$ for the process $e^{+}e^{-}\rightarrow
W^{-} W^{+}\gamma$ is five orders of magnitude more restrictive than
those obtained from the LEP \cite{lep1}. In addition, as shown in
Table III, we improve sensitivity to $\frac{k_{2}^{m}}{\Lambda^{2}}$
coupling with respect to limits derived by Ref. \cite{lhc1}, in
which the best limits on this coupling in the literature are
obtained. An important advantage of the examined $e^{+}e^{-}
\rightarrow e^{+}\gamma^{*} e^{-} \rightarrow e^{+} W^{-} Z \nu_{e}$
process is that it isolates the anomalous $WWZ\gamma$ couplings, and
therefore it enables us to examine $WWZ\gamma$ couplings
independently from $W W \gamma \gamma$ couplings. In Table IV, the
limits on the anomalous couplings $\frac{k_{0}^{W}}{\Lambda^{2}}$
and $\frac{k_{c}^{W}}{\Lambda^{2}}$ are obtained as $[-3.24;\,
3.24]\times 10^{-7}$ and $[-4.71;\, 4.70]\times 10^{-7}$ which can
almost improve the sensitivities up to $37$ times for
$\frac{k_{0}^{W}}{\Lambda^{2}}$ and $\frac{k_{c}^{W}}{\Lambda^{2}}$
with respect to LHC's results. We show in Table V that the best
limits on the anomalous coupling $\frac{a_{n}}{\Lambda^{2}}$ through
the process $e^{+}e^{-} \rightarrow e^{+}\gamma^{*} e^{-}
\rightarrow e^{+} W^{-} Z \nu_{e}$ are calculated as $[-1.17;\,
1.17]\times 10^{-6}$ GeV$^{-2}$ which are more stringent than LEP's
results by almost five orders of magnitude. The best limits on
$\frac{k_{2}^{m}}{\Lambda^{2}}$ via the process $e^{+}e^{-}
\rightarrow e^{+}\gamma^{*} e^{-} \rightarrow e^{+} W^{-} Z \nu_{e}$
are $10$ times than the process $e^{+}e^{-}\rightarrow W^{-}
W^{+}\gamma$ which improves the current experimental limits by a
factor of $1.1$. In addition, we compare our limits with
phenomenological studies on the anomalous couplings
$\frac{k_{2}^{m}}{\Lambda^{2}}$ and $\frac{a_{n}}{\Lambda^{2}}$. Our
limits on $\frac{k_{2}^{m}}{\Lambda^{2}}$ obtained from $e
\gamma^{*}$ collision are 11 times more restrictive than the best
limits obtained with the integrated luminosity of $200$ fb$^{-1}$
corresponding to $W^{\pm}Z\gamma$ production at the $14$ TeV LHC
\cite{lhc1}. These limits are almost of the same order with our
result obtained through the process $e^{+}e^{-}\rightarrow
e^{+}\gamma^{*} e^{-} \rightarrow e^{+} W^{-} Z \nu_{e}$ at the CLIC
with $L_{int}=100$ fb$^{-1}$ and $\sqrt{s}=1.5$ TeV. However, Ref.
\cite{lin4} has considered incoming beam polarizations as well as
the final state polarizations of the gauge bosons in the
cross-section calculations to improve the bounds on anomalous
$\frac{a_{n}}{\Lambda^{2}}$ coupling. We can see that the limits
expected to be obtained for the future $\gamma\gamma$ colliders with
$L_{int}=500$ fb$^{-1}$ and $\sqrt{s}=1.5$ TeV are $5$ times worse
than our best limits when comparing to the unpolarized case. At the
CLIC with $\sqrt{s}=3$ TeV for $L_{int}=590$ fb$^{-1}$, we can set
more stringent limit by two orders of magnitude comparing to the
limits on $\frac{a_{n}}{\Lambda^{2}}$ in Ref.\cite{lhc1}.

We show $95\%$ C.L. contours in the
$\frac{k_{0}^{W}}{\Lambda^{2}}$-$\frac{k_{c}^{W}}{\Lambda^{2}}$
plane for the $e^{+}e^{-}\rightarrow W^{-} W^{+}\gamma$ process in
Figs. $12$-$14$ for various integrated luminosity at $\sqrt{s}=0.5$
, 1 and 3 TeV, respectively. Similarly, the same contours for the
process $e^{+}e^{-} \rightarrow e^{+}\gamma^{*} e^{-} \rightarrow
e^{+} W^{-} Z \nu_{e}$ are depicted in Figs. $15$-$17$. As we can
see from Fig. $14$, the best limits on anomalous couplings
$\frac{k_{0}^{W}}{\Lambda^{2}}$ and $\frac{k_{c}^{W}}{\Lambda^{2}}$
are $[-1.90; 1.92]\times 10^{-7}$ GeV$^{-2}$ and $[-3.34;
3.29]\times 10^{-7}$ GeV$^{-2}$, respectively at $\sqrt{s}=3$ TeV
for $L_{int}=590$ fb$^{-1}$. According to Fig. $17$, the attainable
limits on $\frac{k_{0}^{W}}{\Lambda^{2}}$ and
$\frac{k_{c}^{W}}{\Lambda^{2}}$ are $[-3.86;3.85]\times 10^{-7}$
GeV$^{-2}$ and $[-5.62;5.60]\times 10^{-7}$ GeV$^{-2}$,
respectively.

\section{Conclusions}

The CLIC is an proposed collider with energies on the TeV scale and
extremely high luminosity. Particularly, operating with its high
energy and luminosity is extremely important in order to investigate
geniue anomalous $WWZ\gamma$ quartic gauge couplings that are
described by dimension-six effective Lagrangians. Since energy
dependences of the anomalous couplings are very high, the anomalous
cross sections containing these couplings would have a higher
momentum dependence than the SM cross section. We can easily
understand that the contribution to the cross section of anomalous
quartic couplings rapidly increases when the center-of-mass energy
increases. Moreover, the geniue anomalous couplings can obtain
higher sensitivity via analyzed reactions in the linear colliders
due to very clean experimental conditions and being free from strong
interactions with respect to LHC. Thus in this paper, we have
examined CP-violating and CP-conserving Lagrangians for the
anomalous $WWZ\gamma$ couplings in the processes
$e^{+}e^{-}\rightarrow W^{-} W^{+}\gamma$ and $e^{+}e^{-}
\rightarrow e^{+}\gamma^{*} e^{-} \rightarrow e^{+} W^{-} Z \nu_{e}$
at the CLIC.
\appendix*
\section{The anomalous vertex functions for $W W Z \gamma$}

The anomalous vertex for $W^{+} (p_{1}^{\alpha}) W^{-}(p_{2}^{\beta}) Z (k_{2}^{\nu}) \gamma (k_{1}^{\mu})$ with the help of effective
Lagrangian Eq. ($1$) is generated as follows

\begin{eqnarray}
&&i\frac{\pi\alpha}{4 \textmd{cos}\,\, \theta_{W}\Lambda^{2}}a_{n}[g_{\alpha\nu}[g_{\beta\mu}\, k_{1}.(k_{2}-p_{1})-k_{1\beta}.(k_{2}-p_{1})_{\mu}] \nonumber \\
&&-g_{\beta\nu}[g_{\alpha\mu}\, k_{1}.(k_{2}-p_{2})-k_{1\alpha}.(k_{2}-p_{2})_{\mu}] \nonumber \\ &&+g_{\alpha\beta}[g_{\nu\mu}k_{1}.(p_{1}-p_{2})-k_{1\nu}.(p_{1}-p_{2})_{\mu}] \nonumber \\
&&-k_{2\alpha}(g_{\beta\mu}k_{1\nu}-g_{\nu\mu}k_{1\beta})+k_{2\beta}(g_{\alpha\mu}k_{1\nu}-g_{\nu\mu}k_{1\alpha}) \nonumber \\
&&-p_{2\nu}(g_{\alpha\mu}k_{1\beta}-g_{\beta\mu}k_{1\alpha})+p_{1\nu}(g_{\beta\mu}k_{1\alpha}-g_{\alpha\mu}k_{1\beta}) \nonumber \\
&&+p_{1\beta}(g_{\nu\mu}k_{1\alpha}-g_{\alpha\mu}k_{1\nu})+p_{2\alpha}(g_{\nu\mu}k_{1\beta}-g_{\beta\mu}k_{1\nu})]. \nonumber \\
\end{eqnarray}

In addition, the vertex functions for $W^{+} (p_{1}^{\alpha})
W^{-}(p_{2}^{\beta}) Z (k_{2}^{\nu}) \gamma (k_{1}^{\mu})$ produced
from the effective Lagrangians Eqs. ($8$)-($12$) are expressed below

\begin{eqnarray}
2ie^{2}g^{2}g_{\alpha\beta}[g_{\mu\nu}(k_{1}.k_{2})-k_{1\nu}k_{2\mu}],
\end{eqnarray}

\begin{eqnarray}
&&i\frac{e^{2}g^{2}}{2}[(g_{\mu\alpha}g_{\nu\beta}+g_{\nu\alpha}g_{\mu\beta})(k_{1}.k_{2})+g_{\mu\nu}(k_{2\beta}k_{1\alpha}+k_{1\beta}k_{2\alpha}) \nonumber \\
&&-k_{2\mu}k_{1\alpha}g_{\nu\beta}-k_{2\beta}k_{1\nu}g_{\mu\alpha}-k_{2\alpha}k_{1\nu}g_{\mu\beta}-k_{2\mu}k_{1\beta}g_{\nu\alpha}].
\end{eqnarray}

\begin{eqnarray}
ieg_{z}g^{2}((g_{\mu \alpha}k_{1}.p_{1}-p_{1\mu}k_{1\alpha})g_{\nu
\beta}+(g_{\mu \beta}k_{1}.p_{2} - p_{2\mu}k_{1\beta})g_{\nu
\alpha})
\end{eqnarray}

\begin{eqnarray}
&&i\frac{eg_{z}g^{2}}{2}((k_{1}.p_{1}+k_{1}.p_{2})g_{\mu \nu}g_{\alpha \beta}-(k_{1 \alpha}p_{1\beta}+k_{1\beta}p_{2\alpha})g_{\mu \nu} \nonumber \\
&&-(p_{1\mu}+p_{2\mu})k_{1\nu}g_{\alpha \beta}+(p_{1\beta}g_{\mu\alpha}+p_{2\alpha}g_{\mu \beta})k_{1\nu})
\end{eqnarray}

\begin{eqnarray}
&&i\frac{eg_{z}g^{2}}{2}(k_{1}.p_{1}g_{\mu \beta}g_{\nu \alpha}+k_{1}.p_{2}g_{\mu \alpha}g_{\nu \beta}+(p_{1\nu}-p_{2\nu})k_{1\beta} g_{\mu \alpha} \nonumber \\
&& -(p_{1\nu}-p_{2\nu})k_{1\alpha} g_{\mu \beta}-p_{1\mu}k_{1\beta}g_{\nu \alpha}-p_{2\mu}k_{1\alpha}g_{\nu \beta}).
\end{eqnarray}

\begin{acknowledgements}
This work partially supported by the Abant Izzet Baysal University
Scientific Research Projects under the Project no: 2015.03.02.867.
\end{acknowledgements}

\pagebreak

\begin{figure}
\includegraphics [width=0.8\columnwidth] {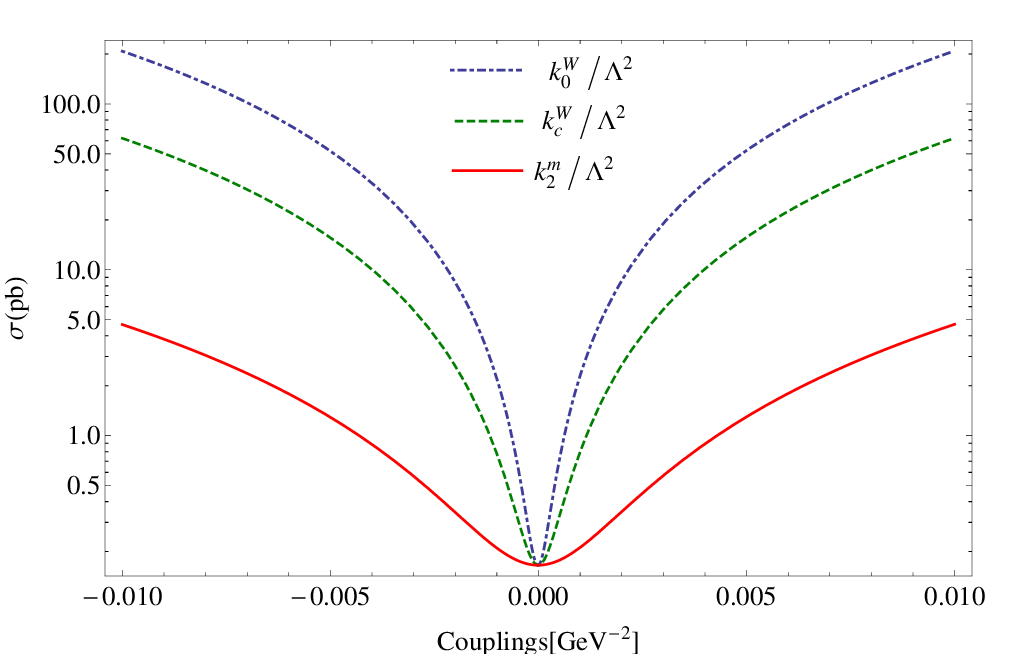}
\caption{The total cross sections as function of anomalous
$\frac{k_{0}^{W}}{\Lambda^{2}}$, $\frac{k_{c}^{W}}{\Lambda^{2}}$ and
$\frac{k_{2}^{m}}{\Lambda^{2}}$ couplings for the
$e^{+}e^{-}\rightarrow W^{-} W^{+}\gamma$ at the CLIC with
$\sqrt{s}=0.5$ TeV. \label{fig1}}
\end{figure}

\begin{figure}
\includegraphics [width=0.8\columnwidth] {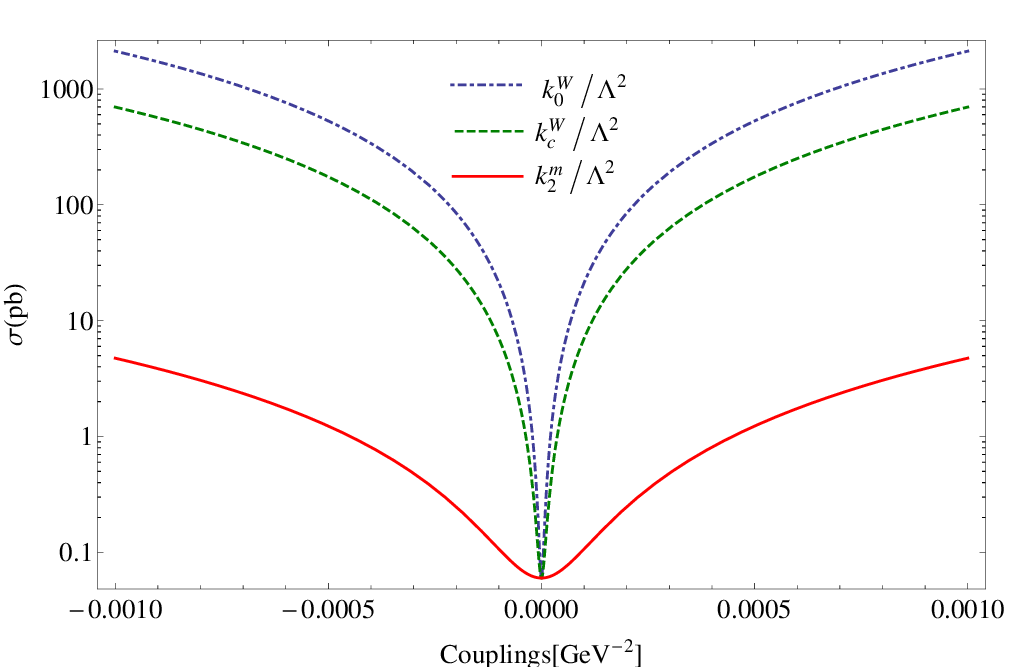}
\caption{The same as Fig. 1 but for $\sqrt{s}=1.5$ TeV.
\label{fig2}}
\end{figure}

\begin{figure}
\includegraphics [width=0.8\columnwidth] {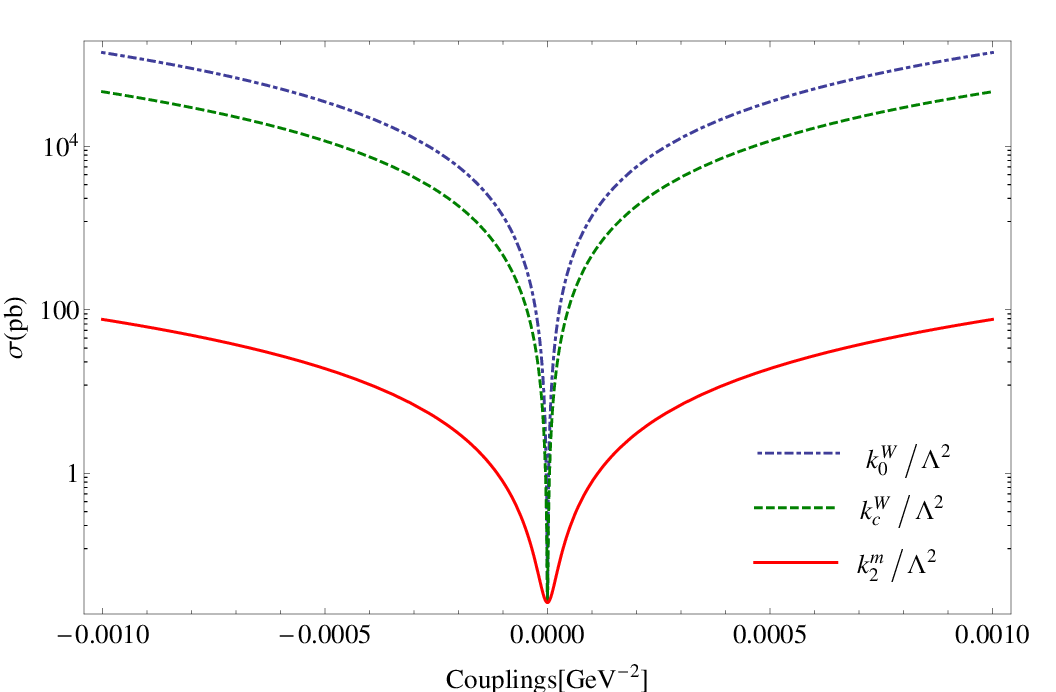}
\caption{The same as Fig. 1 but for $\sqrt{s}=3$ TeV.
\label{fig3}}
\end{figure}

\begin{figure}
\includegraphics [width=0.8\columnwidth] {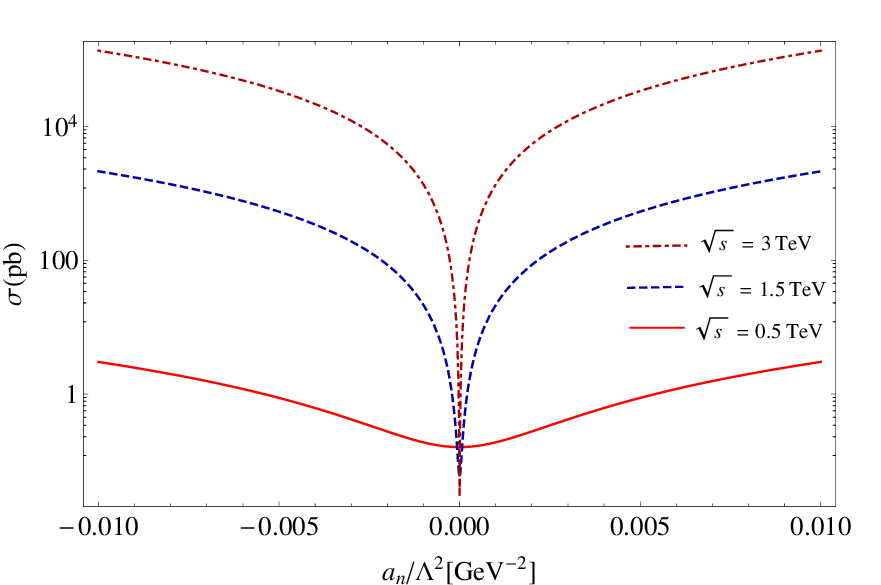}
\caption{The total cross sections as function of anomalous $\frac{a_{n}}{\Lambda^{2}}$ coupling for the process $e^{+}e^{-}\rightarrow W^{-} W^{+}\gamma$ at the CLIC with $\sqrt{s}=0.5,1.5$ and $3$ TeV.
\label{fig4}}
\end{figure}
\begin{figure}
\includegraphics [width=0.8\columnwidth] {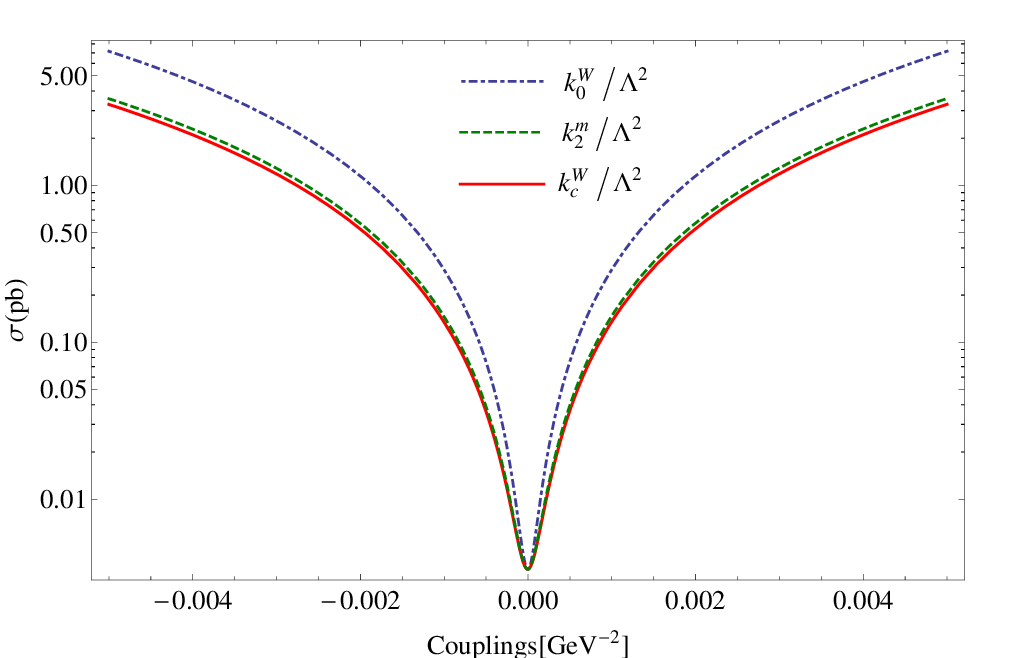}
\caption{The total cross sections as function of anomalous
$\frac{k_{0}^{W}}{\Lambda^{2}}$, $\frac{k_{c}^{W}}{\Lambda^{2}}$ and
$\frac{k_{2}^{m}}{\Lambda^{2}}$ couplings for the process
$e^{+}e^{-} \rightarrow e^{+}\gamma^{*} e^{-} \rightarrow e^{+}
W^{-} Z \nu_{e}$ at the CLIC with $\sqrt{s}=0.5$ TeV. \label{fig5}}
\end{figure}

\begin{figure}
\includegraphics [width=0.8\columnwidth] {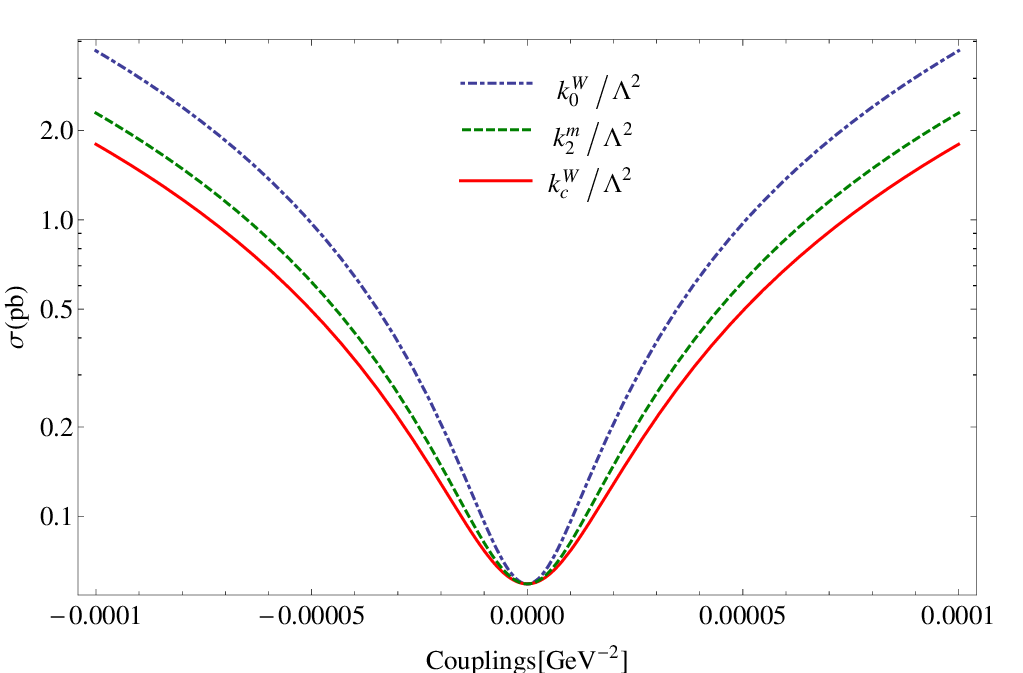}
\caption{The same as Fig. 5 but for $\sqrt{s}=1.5$ TeV.
\label{fig6}}
\end{figure}

\begin{figure}
\includegraphics [width=0.8\columnwidth] {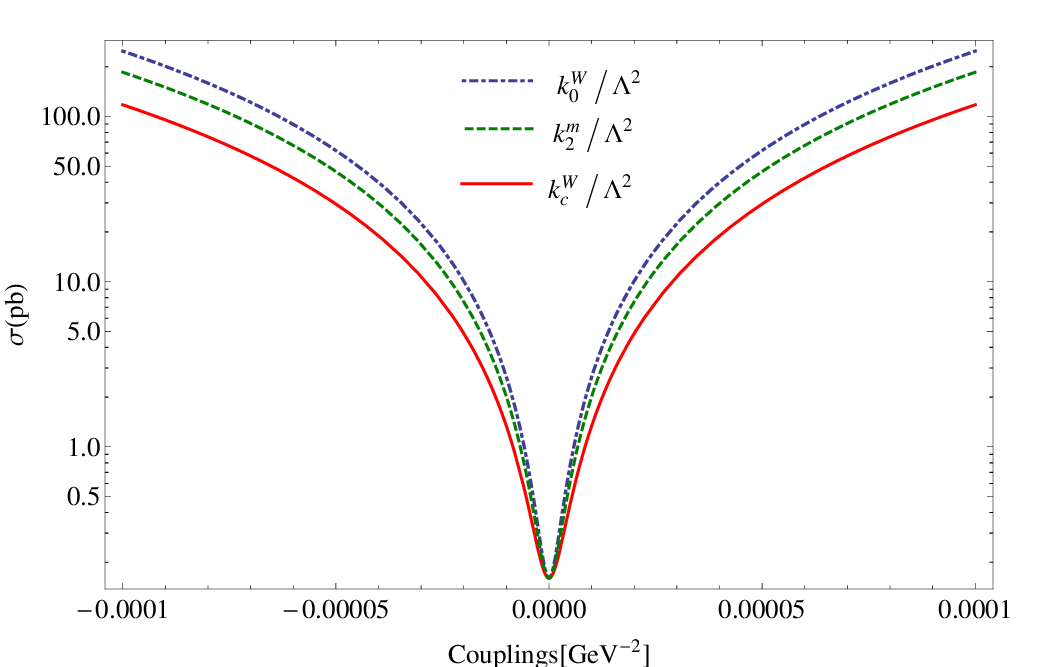}
\caption{The same as Fig. 5 but for $\sqrt{s}=3$ TeV.
\label{fig7}}
\end{figure}

\begin{figure}
\includegraphics [width=0.8\columnwidth] {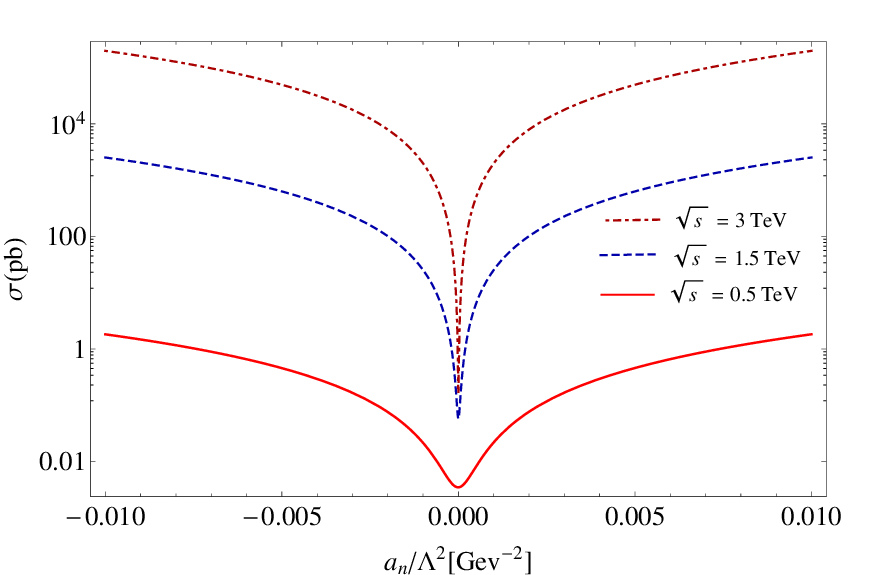}
\caption{The total cross sections as function of anomalous $\frac{a_{n}}{\Lambda^{2}}$ coupling for the process $e^{+}e^{-} \rightarrow e^{+}\gamma^{*} e^{-} \rightarrow e^{+} W^{-} Z \nu_{e}$ at the CLIC with $\sqrt{s}=0.5,1.5$ and $3$ TeV.
\label{fig8}}
\end{figure}

\begin{figure}
\includegraphics {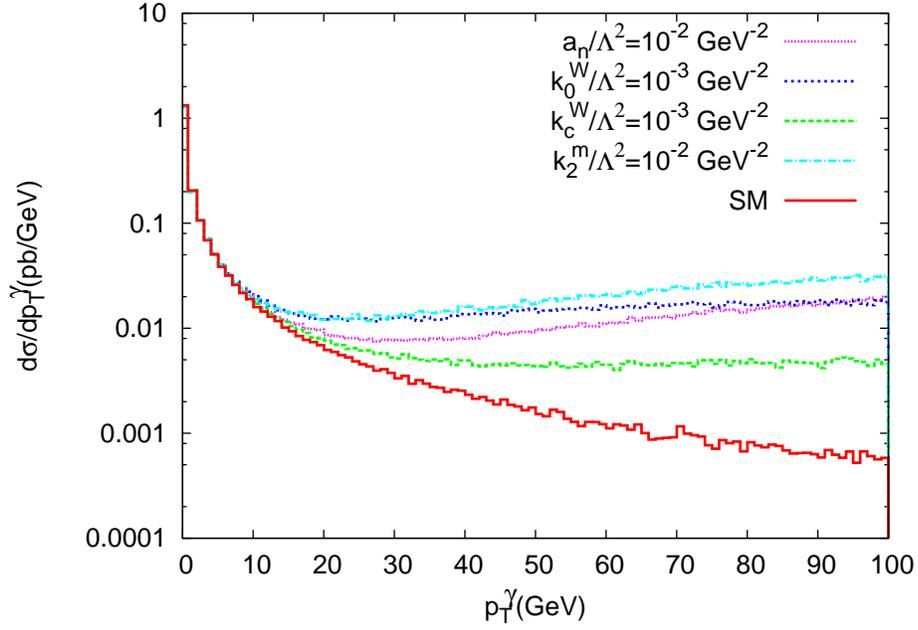}
\caption{The $p_T$ distribution of the final state photon in $e^+e^-\to
W^-W^+\gamma$ process with the anomalous $WWZ\gamma$ couplings
$k_{0,c}^{W}/\Lambda^2$, $k_{2}^{m}/\Lambda^2$ and $a_n/\Lambda^2$ at $\sqrt{s}=0.5$ TeV.
\label{fig9}}
\end{figure}

\begin{figure}
\includegraphics {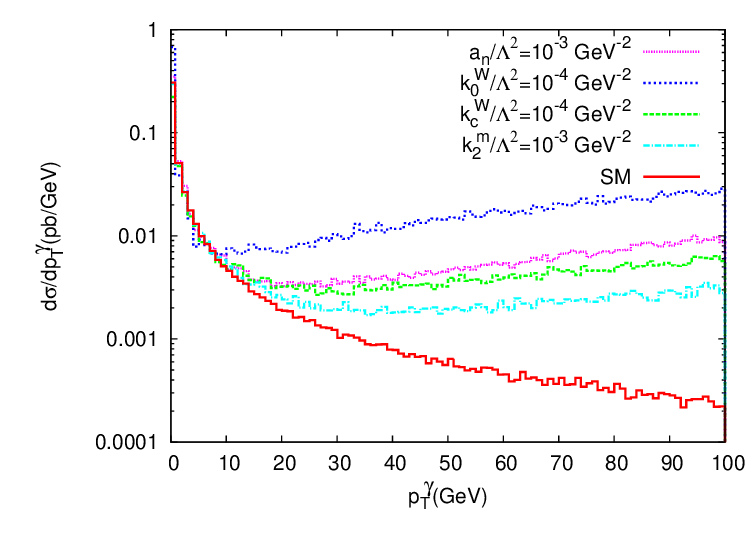}
\caption{The same as Fig. 9 but for $\sqrt{s}=1.5$ TeV.
\label{fig10}}
\end{figure}

\begin{figure}
\includegraphics {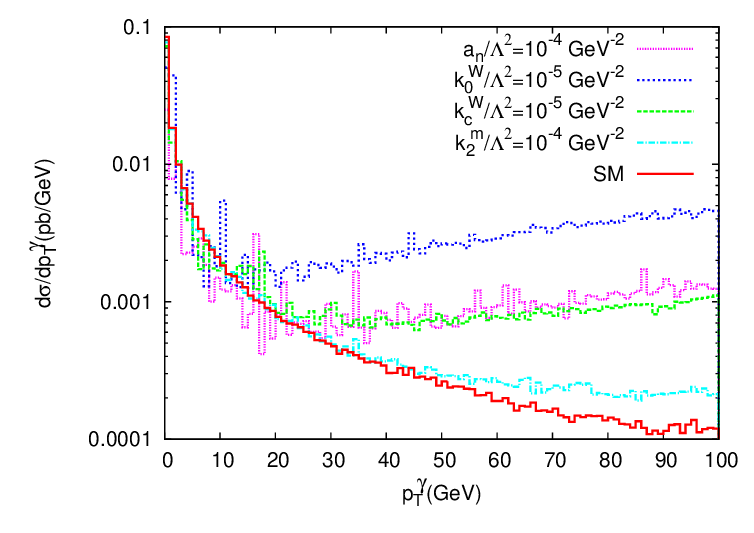}
\caption{The same as Fig. 9 but for $\sqrt{s}=3$ TeV.
\label{fig11}}
\end{figure}

\begin{figure}
\includegraphics {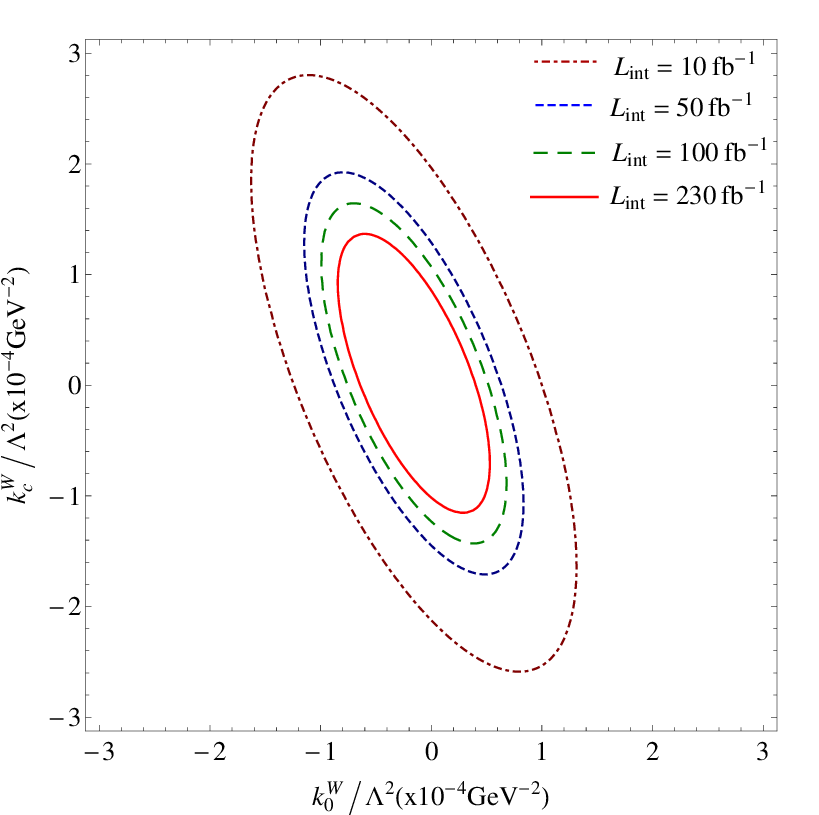}
\caption{$95\%$ C.L. contours for anomalous $\frac{k_{0}^{W}}{\Lambda^{2}}$ and $\frac{k_{c}^{W}}{\Lambda^{2}}$ couplings for the process $e^{+}e^{-}\rightarrow W^{-} W^{+}\gamma$ at the CLIC with $\sqrt{s}=0.5$ TeV.
\label{fig12}}
\end{figure}

\begin{figure}
\includegraphics {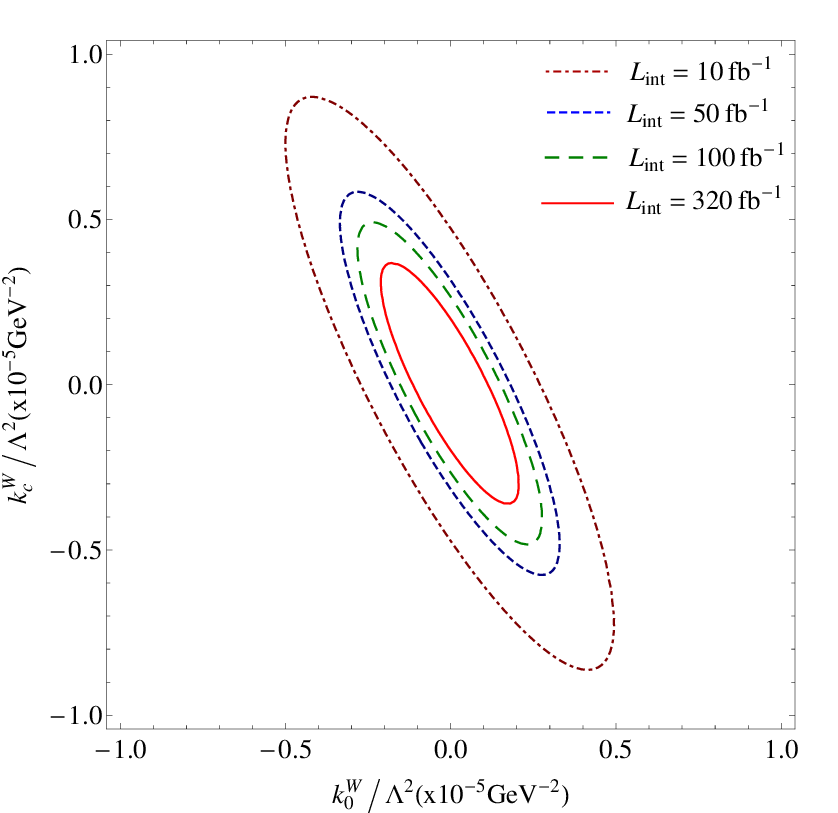}
\caption{The same as Fig. 12 but for $\sqrt{s}=1.5$ TeV.
\label{fig13}}
\end{figure}

\begin{figure}
\includegraphics {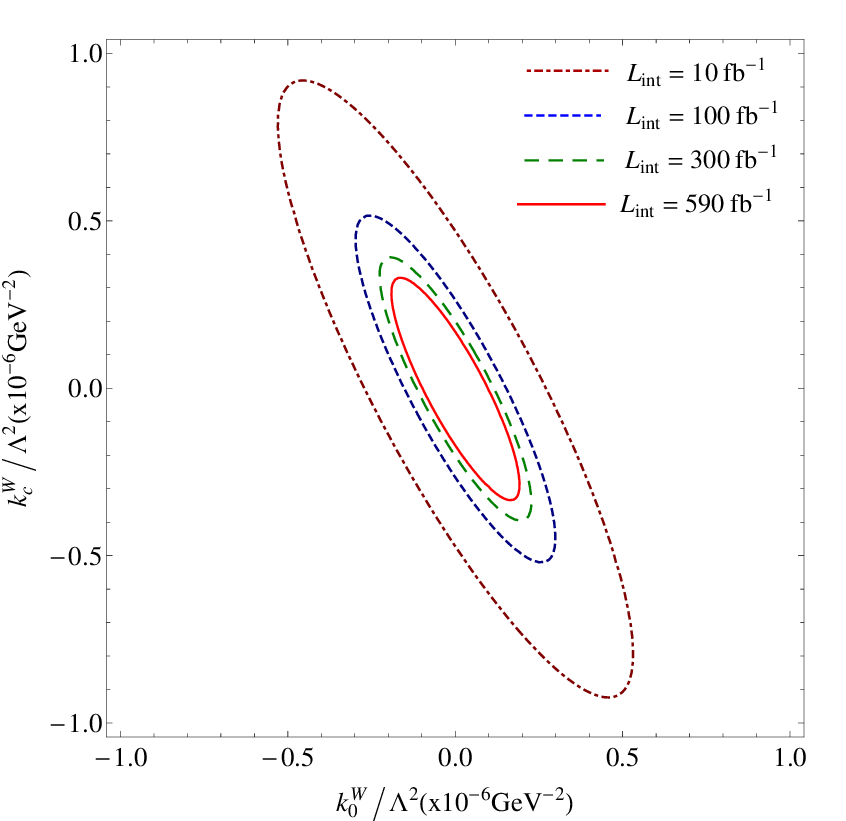}
\caption{The same as Fig. 12 but for $\sqrt{s}=3$ TeV.
\label{fig14}}
\end{figure}

\begin{figure}
\includegraphics {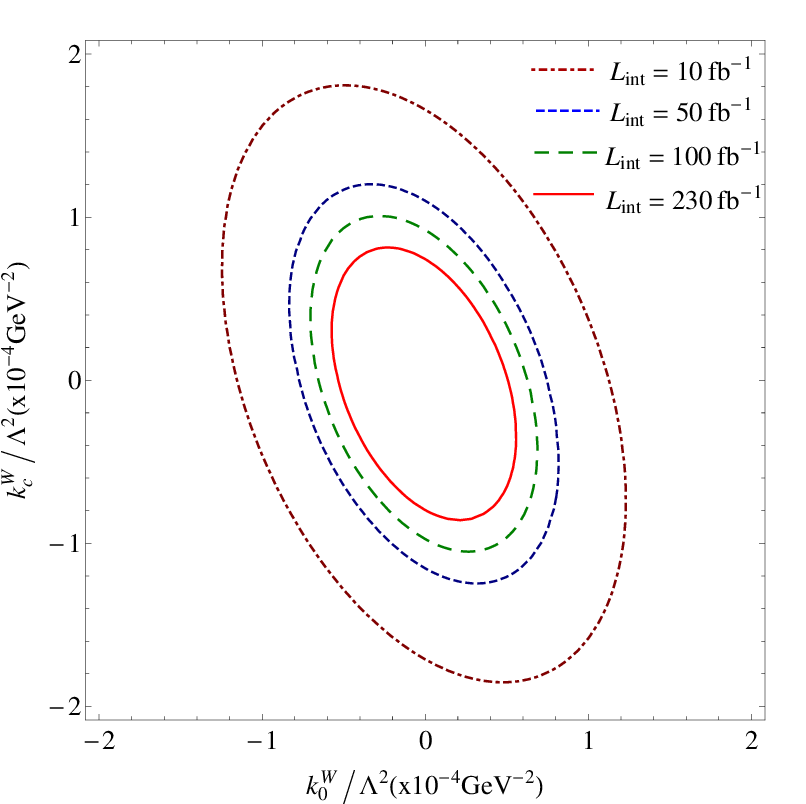}
\caption{$95\%$ C.L. contours for anomalous $\frac{k_{0}^{W}}{\Lambda^{2}}$ and $\frac{k_{c}^{W}}{\Lambda^{2}}$ couplings for the process $e^{+}e^{-} \rightarrow e^{+}\gamma^{*} e^{-} \rightarrow e^{+} W^{-} Z \nu_{e}$ at the CLIC with $\sqrt{s}=0.5$ TeV.
\label{fig15}}
\end{figure}

\begin{figure}
\includegraphics {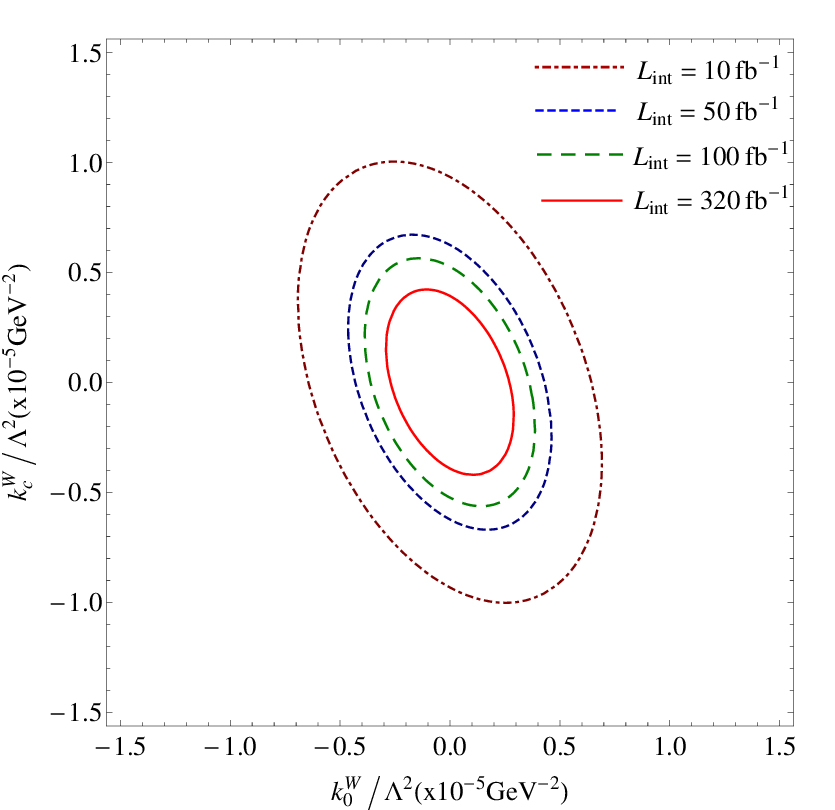}
\caption{The same as Fig. 15 but for $\sqrt{s}=1.5$ TeV.
\label{fig16}}
\end{figure}

\begin{figure}
\includegraphics {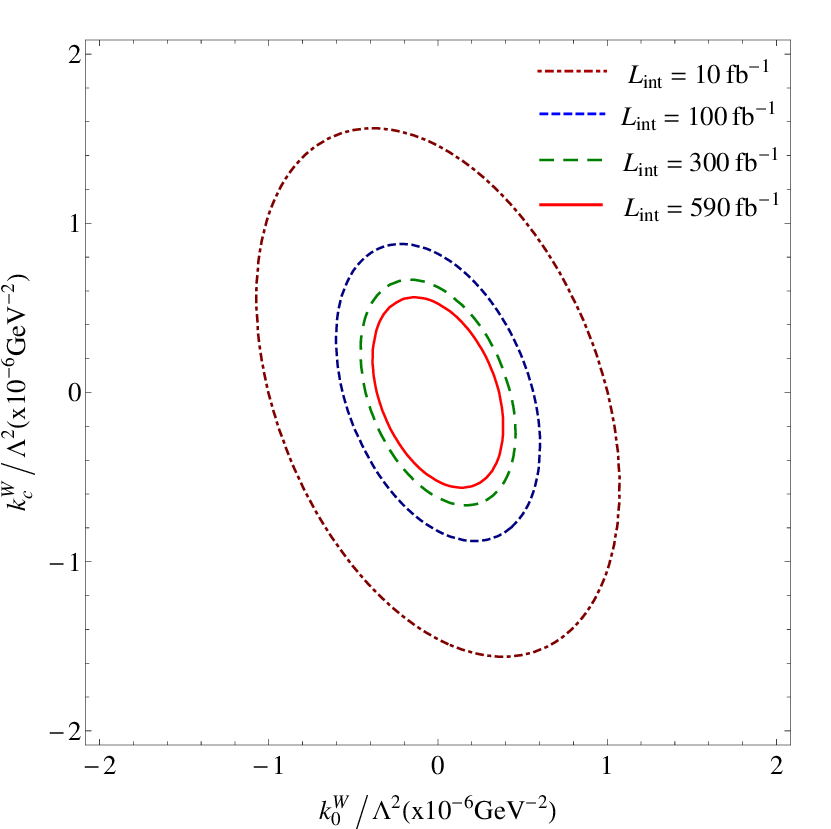}
\caption{The same as Fig. 15 but for $\sqrt{s}=3$ TeV.
\label{fig17}}
\end{figure}

\begin{table}
\caption{$95\%$ C.L. sensitivity bounds of the $\frac{k_{0}^{W}}{\Lambda^{2}}$ and $\frac{k_{c}^{W}}{\Lambda^{2}}$ couplings through the process $e^{+}e^{-}\rightarrow W^{-} W^{+}\gamma$ at the CLIC with $\sqrt{s}=0.5,1.5$ and $3$ TeV.
\label{tab1}}
\begin{ruledtabular}
\begin{tabular} {cccc}
$\sqrt{s}$ (TeV)& $L_{int}$(fb$^{-1}$)& $\frac{k_{0}^{W}}{\Lambda^{2}}$(GeV$^{-2}$)& $\frac{k_{c}^{W}}{\Lambda^{2}}$ (GeV$^{-2}$)\\
\hline
$0.5$& $10$& $[-1.01; 0.99]\times 10^{-4}$& $[-1.83;\, 1.82]\times 10^{-4}$ \\

$0.5$& $50$& $[-6.79;\, 6.50]\times 10^{-5}$& $[-1.22;\, 1.21]\times 10^{-4}$  \\

$0.5$& $100$& $[-5.73;\, 5.50]\times 10^{-5}$& $[-1.03;\, 1.02]\times 10^{-4}$\\

$0.5$& $230$& $[-4.67;\, 4.44]\times 10^{-5}$& $[-8.39;\, 8.32]\times 10^{-5}$ \\
\hline
$1.5$& $10$& $[-2.44;\, 2.43]\times 10^{-6}$& $[-4.24;\, 4.23]\times 10^{-6}$ \\

$1.5$& $50$& $[-1.63;\, 1.61]\times 10^{-6}$& $[-2.83;\, 2.82]\times 10^{-6}$  \\

$1.5$& $100$& $[-1.38;\, 1.36]\times 10^{-6}$& $[-2.38;\, 2.37]\times 10^{-6}$\\

$1.5$& $320$& $[-1.03;\, 1.01]\times 10^{-6}$& $[-1.78;\, 1.77]\times 10^{-6}$ \\
\hline
$3$& $10$& $[-2.43;\, 2.42]\times 10^{-7}$& $[-4.23;\, 4.21]\times 10^{-7}$ \\

$3$& $100$&$[-1.37;\, 1.35]\times 10^{-7}$& $[-2.81;\, 2.79]\times 10^{-7}$  \\

$3$& $300$& $[-1.04;\, 1.03]\times 10^{-7}$& $[-1.81;\, 1.79]\times 10^{-7}$\\

$3$& $590$& $[-8.80;\, 8.73]\times 10^{-8}$& $[-1.53;\, 1.51]\times 10^{-7}$ \\
\end{tabular}
\end{ruledtabular}
\end{table}

\begin{table}
\caption{$95\%$ C.L. sensitivity bounds of the $\frac{a_{n}}{\Lambda^{2}}$ couplings through the process $e^{+}e^{-}\rightarrow W^{-} W^{+}\gamma$ at the CLIC with $\sqrt{s}=0.5,1.5$ and $3$ TeV.
\label{tab2}}
\begin{ruledtabular}
\begin{tabular} {ccc}
$\sqrt{s}$ (TeV)& $L_{int}$(fb$^{-1}$)& $\frac{a_{n}}{\Lambda^{2}}$ (GeV$^{-2}$)\\
\hline
$0.5$& $10$&  $[-8.47;\,8.45]\times 10^{-4}$ \\

$0.5$& $50$&  $[-5.67;\,5.65]\times 10^{-4}$  \\

$0.5$& $100$&  $[-4.77;\,4.75]\times 10^{-4}$\\

$0.5$& $230$&  $[-3.88;\,3.85]\times 10^{-4}$ \\
\hline
$1.5$& $10$&  $[-2.59;\,2.57]\times 10^{-5}$ \\

$1.5$& $50$& $[-1.85;\,1.83]\times 10^{-5}$  \\

$1.5$& $100$&  $[-1.63;\,1.61]\times 10^{-5}$\\

$1.5$& $320$&  $[-1.35;\,1.33]\times 10^{-5}$ \\
\hline
$3$& $10$&  $[-2.46;\,2.46]\times 10^{-6}$ \\

$3$& $100$& $[-1.38;\,1.38]\times 10^{-6}$  \\

$3$& $300$&  $[-1.05;\,1.05]\times 10^{-6}$\\

$3$& $590$&  $[-9.13;\,9.09]\times 10^{-7}$ \\
\end{tabular}
\end{ruledtabular}
\end{table}

\begin{table}
\caption{$95\%$ C.L. sensitivity bounds of the $\frac{k_{2}^{m}}{\Lambda^{2}}$ couplings through the process $e^{+}e^{-}\rightarrow W^{-} W^{+}\gamma$ at the CLIC with $\sqrt{s}=0.5,1.5$ and $3$ TeV.
\label{tab3}}
\begin{ruledtabular}
\begin{tabular} {ccc}
$\sqrt{s}$ (TeV)& $L_{int}$(fb$^{-1}$)& $\frac{k_{2}^{m}}{\Lambda^{2}}$(GeV$^{-2}$)\\
\hline
$0.5$& $10$& $[-6.87;\,6.68]\times 10^{-4}$ \\

$0.5$& $50$&  $[-4.62;\,4.43]\times 10^{-4}$  \\

$0.5$& $100$&  $[-3.90;\,3.72]\times 10^{-4}$ \\

$0.5$& $230$&  $[-3.19;\,3.00]\times 10^{-4}$ \\
\hline
$1.5$& $10$&  $[-5.17;\,5.15]\times 10^{-5}$ \\

$1.5$& $50$&  $[-3.46;\,3.44]\times 10^{-5}$  \\

$1.5$& $100$&  $[-2.91;\,2.89]\times 10^{-5}$\\

$1.5$& $320$&  $[-2.18;\,2.16]\times 10^{-5}$ \\
\hline
$3$& $10$& $[-1.05;\,1.03]\times 10^{-5}$ \\

$3$& $100$& $[-5.92;\,5.79]\times 10^{-6}$  \\

$3$& $300$& $[-4.51;\,4.38]\times 10^{-6}$ \\

$3$& $590$& $[-3.82;\,3.69]\times 10^{-6}$ \\
\end{tabular}
\end{ruledtabular}
\end{table}

\begin{table}
\caption{$95\%$ C.L. sensitivity bounds of the $\frac{k_{0}^{W}}{\Lambda^{2}}$ and $\frac{k_{c}^{W}}{\Lambda^{2}}$ couplings through the processes $e^{+}e^{-} \rightarrow e^{+}\gamma^{*} e^{-} \rightarrow e^{+} W^{-} Z \nu_{e}$ at the  CLIC with $\sqrt{s}=0.5,1.5$ and $3$ TeV.
\label{tab4}}
\begin{ruledtabular}
\begin{tabular}{cccc}
$\sqrt{s}$ (TeV)& $L_{int}$(fb$^{-1}$)& $\frac{k_{0}^{W}}{\Lambda^{2}}$(GeV$^{-2}$)& $\frac{k_{c}^{W}}{\Lambda^{2}}$ (GeV$^{-2}$)\\
\hline
$0.5$& $10$& $[-1.03;\, 1.01]\times 10^{-4}$& $[-1.53;\, 1.48]\times 10^{-4}$ \\

$0.5$& $50$& $[-6.97;\, 6.69]\times 10^{-5}$& $[-1.04;\, 0.98]\times 10^{-4}$ \\

$0.5$& $100$& $[-5.88;\, 5.60]\times 10^{-5}$& $[-8.75;\, 8.22]\times 10^{-5}$\\

$0.5$& $230$& $[-4.80;\, 4.52]\times 10^{-5}$& $[-7.16;\, 6.62]\times 10^{-5}$ \\
\hline
$1.5$& $10$& $[-5.76;\, 5.75]\times 10^{-6}$& $[-8.37;\, 8.35]\times 10^{-6}$ \\

$1.5$& $50$& $[-3.86;\, 3.85]\times 10^{-6}$& $[-5.60;\, 5.58]\times 10^{-6}$ \\

$1.5$& $100$& $[-3.24;\, 3.23]\times 10^{-6}$& $[-4.71;\, 4.69]\times 10^{-6}$\\

$1.5$& $320$& $[-2.43;\, 2.42]\times 10^{-6}$& $[-3.53;\,3.50]\times 10^{-6}$ \\
\hline
$3$& $10$& $[-8.98;\, 8.97]\times 10^{-7}$& $[-1.31;\, 1.30]\times 10^{-6}$ \\

$3$& $100$&$[-5.05;\, 5.04]\times 10^{-7}$& $[-7.34;\, 7.33]\times 10^{-7}$  \\

$3$& $300$& $[-3.84;\, 3.83]\times 10^{-7}$& $[-5.58;\, 5.57]\times 10^{-7}$\\

$3$& $590$& $[-3.24;\, 3.24]\times 10^{-7}$& $[-4.71;\, 4.70]\times 10^{-7}$ \\
\end{tabular}
\end{ruledtabular}
\end{table}

\begin{table}
\caption{$95\%$ C.L. sensitivity bounds of the $\frac{a_{n}}{\Lambda^{2}}$ couplings through the processes $e^{+}e^{-} \rightarrow e^{+}\gamma^{*} e^{-} \rightarrow e^{+} W^{-} Z \nu_{e}$ at the  CLIC with $\sqrt{s}=0.5,1.5$ and $3$ TeV.
\label{tab5}}
\begin{ruledtabular}
\begin{tabular}{ccc}
$\sqrt{s}$ (TeV)& $L_{int}$(fb$^{-1}$)& $\frac{a_{n}}{\Lambda^{2}}$ (GeV$^{-2}$)\\
\hline
$0.5$& $10$&  $[-4.08;\, 3.96]\times 10^{-4}$ \\

$0.5$& $50$&  $[-2.75;\, 2.63]\times 10^{-4}$  \\

$0.5$& $100$&  $[-2.33;\, 2.20]\times 10^{-4}$\\

$0.5$& $230$& $[-1.90;\, 1.78]\times 10^{-4}$ \\
\hline
$1.5$& $10$&  $[-2.19;\, 2.17]\times 10^{-5}$ \\

$1.5$& $50$&  $[-1.47;\, 1.45]\times 10^{-5}$  \\

$1.5$& $100$&  $[-1.23;\, 1.22]\times 10^{-5}$\\

$1.5$& $320$&  $[-9.26;\, 9.07]\times 10^{-6}$ \\
\hline
$3$& $10$&  $[-3.16;\, 3.16]\times 10^{-6}$ \\

$3$& $100$& $[-1.78;\, 1.77]\times 10^{-6}$  \\

$3$& $300$&  $[-1.35;\, 1.35]\times 10^{-6}$\\

$3$& $590$& $[-1.17;\, 1.17]\times 10^{-6}$ \\
\end{tabular}
\end{ruledtabular}
\end{table}

\begin{table}
\caption{$95\%$ C.L. sensitivity bounds of the $\frac{k_{2}^{m}}{\Lambda^{2}}$ couplings through the processes $e^{+}e^{-} \rightarrow e^{+}\gamma^{*} e^{-} \rightarrow e^{+} W^{-} Z \nu_{e}$ at the  CLIC with $\sqrt{s}=0.5,1.5$ and $3$ TeV.
\label{tab6}}
\begin{ruledtabular}
\begin{tabular}{ccc}
$\sqrt{s}$ (TeV)& $L_{int}$(fb$^{-1}$)& $\frac{k_{2}^{m}}{\Lambda^{2}}$(GeV$^{-2}$)\\
\hline
$0.5$& $10$&  $[-1.48;\, 1.41]\times 10^{-4}$ \\

$0.5$& $50$& $[-9.98;\, 9.36]\times 10^{-5}$  \\

$0.5$& $100$& $[-8.45;\, 7.42]\times 10^{-5}$\\

$0.5$& $230$& $[-6.92;\, 6.29]\times 10^{-5}$ \\
\hline
$1.5$& $10$& $[-7.38;\, 7.37]\times 10^{-6}$ \\

$1.5$& $50$& $[-4.94;\, 4.92]\times 10^{-6}$  \\

$1.5$& $100$& $[-4.15;\, 4.14]\times 10^{-6}$\\

$1.5$& $320$& $[-3.11;\, 3.09]\times 10^{-6}$\\
\hline
$3$& $10$&  $[-1.04;\, 1.04]\times 10^{-6}$\\

$3$& $100$&$[-5.85;\, 5.84]\times 10^{-7}$  \\

$3$& $300$& $[-4.44;\, 4.43]\times 10^{-7}$\\

$3$& $590$& $[-3.75;\, 3.74]\times 10^{-7}$\\
\end{tabular}
\end{ruledtabular}
\end{table}


\begin{thebibliography}{99}
\bibitem{higgs1} S. Chatrchyan {\it et al.}, CMS Collaboration, Phys. Lett. B 716, 30 (2012).
\bibitem{higgs2} G. Aad {\it et al.}ATLAS Collaboration, Phys. Lett. B 716, 1 (2012).
\bibitem{lag1} O. J. P. Eboli, M. C. Gonzalez-Garcia and S. F. Novaes, Nucl. Phys. B411, 381 (1994).
\bibitem{lhc} O. J. P. Eboli, M.C. Gonzalez-Garcia, S. M. Lietti, Phys. Rev. D 69, 095005 (2004).
\bibitem{lag3} G. Belanger, F. Boudjema, Y. Kurihara, D. Perret-Gallix, A. Semenov, Eur. Phys. J. C 13, 283 (2000).
\bibitem{lhc1} K.~Ye, D.~Yang and Q.~Li, Phys. Rev. D 88, 015023
(2013).
\bibitem{lin} G. Abu Leil and W. J. Stirling, J. Phys. G 21, 517 (1995).
\bibitem{linb} W. J. Stirling and A. Werthenbach, Eur. Phys. J. C14, 103 (2000).
\bibitem{linc} A. Denner {\it et al.}, Eur. Phys. J. C 20, 201 (2001).
\bibitem{lind} G. Montagna {\it et al.}, Phys. Lett. B 515, 197 (2001).
\bibitem{line} M. Beyer {\it et al.}, Eur. Phys. J. C 48, 353 (2006).
\bibitem{lin2} I. Sahin, J. Phys. G: Nucl. Part. Phys. 35, 035006 (2008).
\bibitem{lin3} O.~J.~P.~Eboli, M.~B.~Magro, P.~G.~Mercadante and S.~F.~Novaes, Phys. Rev. D 52, 15 (1995).
\bibitem{lin4} I. Sahin, J. Phys.  G: Nucl. Part. Phys. 36, 075007 (2009).
\bibitem{lep1} P. Achard {\it et al.}, L3 Collaboration, Phys. Lett. B 527, 29 (2002).
\bibitem{lep2} J. Abdallah {\it et al.}, DELPHI Collaboration, Eur. Phys. J. C31, 139 (2003).
\bibitem{lep3} G. Abbiendi {\it et al.}, OPAL Collaboration, Phys. Lett. B580, 17 (2004).
\bibitem{s覺n覺r}  S.~Chatrchyan {\it et al.}  [CMS Collaboration], Phys. Rev. D 90, 032008
(2014).
\bibitem{clic} D. Dannheim {\it et al.}, CLIC $e^{+}e^{-}$ Linear Collider Studies, arXiv:1305.5766.
\bibitem{clic1} D. Dannheim {\it et al.}, CLIC $e^{+}e^{-}$ Linear Collider Studies, arXiv:1208.1402.
\bibitem{clic2} L. Linssen, A. Miyamoto, M. Stanitzki and H. Weerts, CERN-2012-003 ; ANL-HEP-TR-12-01 ; DESY-12-008 ; KEK-Report-2011-7.
\bibitem{las1} I. F. Ginzburg, G. L. Kotkin, V. G. Serbo and V. I. Telnov, Nucl. Instr. and Meth. 205, 47 (1983).
\bibitem{las2} I. F. Ginzburg, G. L. Kotkin, S. L. Panfil, V. G. Serbo and V. I. Telnov, Nucl. Instr. and Meth. 219, 5 (1984).
\bibitem{Brodsky:1971ud} S.~J.~Brodsky, T.~Kinoshita and H.~Terazawa, Phys. Rev. D 4, 1532 (1971).
\bibitem{Terazawa:1973tb} H.~Terazawa, Rev. Mod. Phys. 45, 615 (1973).
\bibitem{es1} V.M. Budnev, I.F. Ginzburg, G.V. Meledin and V.G. Serbo, Phys. Rept. 15, 181 (1974).
\bibitem{es2} K. Piotrzkowski, Phys. Rev. D 63, 071502 (2001).
\bibitem{es3} G. Baur et al., Phys. Rep. 364, 359 (2002).
\bibitem{a1} J. Abdallah {\it et al.}, DELPHI Collaboration, Eur. Phys. J. C 35, 159 (2004).
\bibitem{a2} A. Abulencia {\it et al.}, CDF Collaboration, Phys. Rev. Lett. 98, 112001 (2007).
\bibitem{a3} T. Aaltonen {\it et al.}, CDF Collaboration, Phys. Rev. Lett. 102, 222002 (2009).
\bibitem{a4} T. Aaltonen {\it et al.}, CDF Collaboration, Phys. Rev. Lett. 102, 242001 (2009).
\bibitem{Abazov:2010bk} V.~M.~Abazov {\it et al.}  [D0 Collaboration], Phys. Rev. D 88 012005 (2013).
\bibitem{Tasevsky:2011zz} M.~Tasevsky [ATLAS Collaboration], AIP Conf. Proc. 1350, 164 (2011).
\bibitem{a5} S. Chatrchyan {\it et al.}, CMS Collaboration, JHEP 1201, 052 (2012).
\bibitem{a6} S. Chatrchyan et al:, CMS Collaboration, JHEP 1211, 080 (2012).
\bibitem{a7} S. Atag and A. Billur, JHEP 11, 060 (2010).
\bibitem{a8} S. Atag, S. C. \.{I}nan and \.{I}. \c{S}ahin, Phys. Rev. D 80, 075009 (2009).
\bibitem{a9} \.{I}. \c{S}ahin and S. C. \.{I}nan, JHEP 09, 069 (2009).
\bibitem{a10} S. C. \.{I}nan, Phys. Rev. D 81, 115002 (2010).
\bibitem{a11} \.{I}. \c{S}ahin and M. K\"{o}ksal, JHEP 11, 100 (2011).
\bibitem{a12} M. K\"{o}ksal and S. C. \.{I}nan, Advances in High Energy Physics, Volume 2014, Article ID 935840, 11 Pages (2014).
\bibitem{a13} M. K\"{o}ksal and S. C. \.{I}nan, Advances in High Energy Physics Volume 2014, Article ID 315826, 8 pages (2014).
\bibitem{a14} A. A. Billur and M. K\"{o}ksal, Phys. Rev. D 89, 037301 (2014).
\bibitem{a15} A. A. Billur and M. K\"{o}ksal, arXiv:1311.5326.
\bibitem{a16} A. Senol, Phys. Rev. D 87, 073003 (2013).
\bibitem{a17} A.~Senol, Int.\ J.\ Mod.\ Phys.\ A 29, 1450148 (2014).
\bibitem{a18} I.~Sahin, A.~A.~Billur, S.~C.~Inan, B.~Sahin, M.~K闥sal, P.~Tektas, E.~Alici and R.~Yildirim, Phys. Rev. D 88, 095016
(2013).
\bibitem{a19} S. C. \.{I}nan and A. Billur, Phys. Rev. D 84, 095002 (2011).
\bibitem{a20} \.{I}. \c{S}ahin, Phys. Rev. D 85, 033002 (2012).
\bibitem{a21} \.{I}. \c{S}ahin and B. \c{S}ahin, Phys. Rev. D 86, 115001 (2012).
\bibitem{a22} B. \c{S}ahin and A. A. Billur, Phys. Rev. D 86, 074026 (2012).
\bibitem{a23} A. A. Billur, Europhys. Lett. 101, 21001 (2013).
\bibitem{a24} M. Tasevsky, Nucl. Phys. Proc. Suppl. 179-180, 187 (2008).
\bibitem{a25} M. Tasevsky, arXiv:0910.5205.
\bibitem{a26} H. Sun, Nucl. Phys. B 886, 691 (2014).
\bibitem{a27} H. Sun and Chong-Xing Yue, Eur. Phys. J. C 74, 2823 (2014).
\bibitem{comphep} A. Pukhov {\it et al.}, Report No. INP MSU 98-41/542; arXiv:hep-ph/9908288; arXiv:hep-ph/0412191.

\end{thebibliography}
\end{document}